\documentstyle[12pt]{article}
\begin{document}

\tolerance=5000

\def\cL{{\cal L}}
\def\be{\begin{equation}}
\def\ee{\end{equation}}
\def\bea{\begin{eqnarray}}
\def\eea{\end{eqnarray}}
\def\tr{{\rm tr}\, }
\def\nn{\nonumber \\}
\def\gd{g^\dagger}
\def\e{{\rm e}}

\begin{flushright}
NDA-FP-25 \\
March 1996 \\
hep-th/yymmxxx \\
\end{flushright}

\vfill

\begin{center}

{\Large\bf
Is the Chiral Model equivalent to
Wess-Zumino-Witten Model when coupled 
with Gravity?
}

\vfill

{\Large\sc Shin'ichi NOJIRI}

\vfill

{\large\sl Department of Mathematics and Physics \\
National Defence Academy \\
Hashirimizu Yokosuka 239, JAPAN}

\vfill

{\bf ABSTRACT}
\end{center}

We investigate the non-abelian $T$-duality of 
Wess-Zumino-Witten model. The obtained dual model is
equivalent to the model dual to the $SU(2)$ chiral model 
found by Curtright-Zachos.
This might tell that the Wess-Zumino term would be induced 
when the chiral model couples with gravity.

\newpage

When we quantize the gravity in two dimensions, we usually 
choose a conformal gauge \cite{polyakov}. 
Since the residual gauge symmetry 
is conformal symmetry, the gauge fixed system should be 
described by a conformal field theory with the vanishing 
central charge. 
An interesting question is what happens 
when a massive system is coupled with the gravity.
In this brief paper, we consider the $SU(2)$ chiral model, 
which is known to have a mass gap \cite{polywieg} 
so that it does not have 
conformal symmetry. 
%When the system couples with the gravity, 
%the system where the conformal gauge is chosen should has 
%conformal symmetry. 
Since the conformal invariant system 
having the same global symmetry as the chiral model 
is $SU(2)$ Wess-Zumino-Witten model \cite{WZW}, 
it might be natural to consider 
that the Wess-Zumino term would be induced when 
the chiral model couples with gravity.
In order to investigate this conjecture, we use the 
non-abelian $T$-duality \cite{quevedo}. 
When a $\sigma$ model 
in two dimensions has an abelian or non-abelian isometry, 
we can introduce the gauge fields 
by gauging the isometry and impose a constraint which  
makes the gauge curvature vanish. 
After imposing a gauge fixing condition and integrating 
gauge fields, we obtain the dual model.
In this model, there appears a 
shift of the dilaton fields when we use a 
regularization which keeps the general covariance. 
In this paper, we show that a model dual to 
the chiral model is identical with the model 
dual to the Wess-Zumino-Witten model. 
At first sight, it appears that there would be 
a discrepancy since the chiral model has a mass gap 
although the Wess-Zumino-Model is a massless theory. 
The key to solve this discrepancy would be 
the dilaton term. 
In the Wess-Zumino-Witten model, the dilaton term is 
necessary in order to keep the conformal symmetry.
On the other hand, if we consider the chiral model in the 
flat space-time and do not use the regularization which keeps 
the general covariance, there does not appear 
the dilaton term. 
Curtright and Zachos have shown that the dual model 
without the dilaton term is equivalent to the chiral 
model by using canonical transformation \cite{zachos}.
This suggests that the chiral model is equivalent to the 
Wess-Zumino-Witten model only when coupled with gravity.

The action of $SU(2)$ chiral model is given by
\be
\label{chilag}
\cL=-{1 \over e^2}\tr \partial \gd \bar\partial g \ .
\ee
Here $g$ is a group element of $SU(2)$ ($g^{-1}=\gd$) 
and $e$ is a coupling constant.
The Lagrangean (\ref{chilag}) is invariant under the left and 
right $SU(2)$ transformation,
\bea
&&g\rightarrow hg\ \ \ ({\rm left})\ ,\ \ \ \ 
g\rightarrow gh\ \ \ ({\rm right}) \ .
\eea
($h$ is a group element of $SU(2)$.)
In order to obtain a dual Lagrangean, we gauge left or right 
$SU(2)$ symmetry by introducing gauge fields $A$ and $\bar A$,
\bea
\label{gaugef}
&&A=A^iT^i\ ,\ \ \ \bar A=\bar A^iT^i \\
&&T^i={1 \over 2}\sigma^i\ , 
\eea
($\sigma^i$'s are Pauli matrices.)
\bea
\label{gauged}
\cL^{\rm L}&=&{1 \over e^2}\tr \gd (\partial g+ A g)
\gd (\bar\partial g + \bar A g)\ \ \ ({\rm left}) \nn
\cL^{\rm R}&=&{1 \over e^2}\tr \gd (\partial g+ g A)
\gd (\bar\partial g + g \bar A )\ \ \ ({\rm right}) 
\eea
and add a term which makes the gauge curvature 
$F=[\partial + A, \bar\partial + \bar A]$ vanish,
\be
\label{constraint1}
\cL^{\rm constraint}={2 \over e^2}\tr \theta F
\ee
Here $\theta$ is an element of $SU(2)$ algebra
\be
\label{constr}
\theta=\theta^i T^i\ .
\ee
When we integrate the gauge fields $A$ and $\bar A$ by 
choosing the gauge condition
\be
\label{condition}
g=1\ ,
\ee
we obtain the dual Lagrangean
\be
\label{dualchi}
\cL^{\rm dual}={1 \over e^2}{4 \over 1 + 4 (\theta)^2}
\left[\delta_{ij}-2\epsilon_{ijk}\theta^k +4\theta^i\theta^j
\right]\partial\theta^i\bar\partial\theta^j \ .
\ee
Here $(\theta)^2=\theta^i\theta^i$.
The Lagrangean (\ref{dualchi}) was found by Curtright and Zachos 
\cite{zachos}, 
who proved the equivalence between the two Lagrangeans (\ref{chilag}) 
and (\ref{dualchi}) by using canonical transformation. 

When we couple the system with the gravity
and use the regularization which keeps 
the general covariance, there appears a dilaton 
term in the dual theory \cite{quevedo},
\be
\label{dilaton}
\cL^{{\rm dilaton}}=
-{1 \over 4\pi}R^{(2)}\ln (1 + 4 (\theta)^2)\ .
\ee

In the following, we consider the $SU(2)$ Wess-Zumino-Witten model 
whose action is given by,
\bea
\label{WZW}
S&=&-{k \over 4\pi}\int_{\partial B}d^2 \tr \left(
\partial \gd \bar\partial g \right) \nn
&-&{k \over 2\pi}\int_B dV \epsilon^{\mu\nu\rho}
\tr\left(\gd\partial_\mu g\gd\partial_\nu g\gd \partial_\rho 
g\right)
\eea
Here $k$ is an integer parameter called ``level'' and $B$ is a 
three dimensional manifolds whose boundary $\partial B$ is a 
two dimensional (string) world sheet.
When we parametrize the $SU(2)$ group element $g$ by the Euler 
angles
\be
\label{Euler}
g=\e^{{i \over 2}\phi_L\sigma_2}\e^{{i \over 2}r\sigma_1}
\e^{{i \over 2}\phi_R\sigma_2}\ ,
\ee
the action (\ref{WZW}) can be given by a local Lagrangean
\be
\label{LWZW}
\cL^{{\rm WZW}}={k \over 4\pi}
\left\{
\partial\phi_L\bar\partial\phi_L
+\partial\phi_R\bar\partial\phi_R
+2\cos r \partial\phi_L\bar\partial\phi_R
+\partial r \bar \partial r
\right\} \ .
\ee
If we define new variables by
\be
\label{newvari}
l\equiv {r \over 2}\ , \ \ \ \varphi^\pm\equiv
{\phi_L \pm \phi_R \over 2} \ ,
\ee
the infinitesimal right(left)-handed $SU(2)$ transformation 
$\delta_L g={i \over 2}\epsilon_L^i T^i g$
($\delta_R g={i \over 2}g\epsilon_R^i T^i$) is given in terms
of these new variables as follows
\bea
\label{LRtrsf}
\delta l &=& {1 \over 2}\Bigl\{
\epsilon_L^1(\sin \varphi^+ \sin \varphi^- 
+ \cos^2\varphi^-) 
+\epsilon_L^3(\sin \varphi^+ \cos \varphi^- 
- \cos\varphi^+ \sin \varphi^-) \nn
&& + \epsilon_R^1(\sin \varphi^+ \sin \varphi^- 
- \cos^2\varphi^-) 
+\epsilon_R^3(\sin \varphi^+ \cos \varphi^- 
+ \cos\varphi^+ \sin \varphi^-)
\Bigr\} \nn
&\equiv& a^i_L \epsilon^i_L + a^i_R \epsilon^i_R \nn
\delta \varphi^+ &=& {1 \over 2}\Bigl\{
\epsilon_L^1 \tan l (-\cos \varphi^+ \sin \varphi^- 
+\tan \varphi^+ \cos^2 \varphi^-)
+\epsilon_L^2 \nn
&& +\epsilon_L^3 \tan l (-\cos\varphi^+ \cos \varphi^- 
- \sin \varphi^+ \sin \varphi^-) \nn
&& +\epsilon_R^1 \tan l (-\cos \varphi^+ \sin \varphi^- 
-\tan \varphi^+ \cos^2 \varphi^-)
-\epsilon_R^2 \nn
&& +\epsilon_R^3 \tan l (-\cos\varphi^+ \cos \varphi^- 
+ \sin \varphi^+ \sin \varphi^-) \Bigr\} \nn
&\equiv& b^{+i}_L \epsilon^i_L + b^{+i}_R \epsilon^i_R \nn
\delta \varphi^- &=& {1 \over 2}\Bigl\{
\epsilon_L^1 \cot l (\sin \varphi^+ \cos \varphi^- 
-\sin \varphi^- \cos \varphi^-)
-\epsilon_L^2 \nn
&& +\epsilon_L^3 \cot l ( -\sin \varphi^+ \sin \varphi^-
- \cos \varphi^+ \cos \varphi^-) \nn
&&+\epsilon_R^1 \cot l (\sin \varphi^+ \cos \varphi^- 
+\sin \varphi^- \cos \varphi^-)
+\epsilon_R^2 \nn
&&+\epsilon_R^3 \cot l ( -\sin \varphi^+ \sin \varphi^-
+ \cos \varphi^+ \cos \varphi^-) \Bigr\} \nn
&\equiv& b^{-i}_L \epsilon^i_L + b^{-i}_R \epsilon^i_R 
\eea
In order to obtain the dual theory, 
left-handed or right-handed 
$SU(2)$ symmetry should be gauged. 
The gauged action is given by
using the notation of Eq.(\ref{LRtrsf})
\bea
\label{gLWZW}
\cL^{{\rm WZW}}&=&{k \over 4\pi}
\Bigl[
2(\partial\varphi^+ -b^{+i}_{L,R}A^i_{L,R})
(\bar\partial\varphi^+ -b^{+i}_{L,R}\bar A^i_{L,R}) \nn
&& +2(\partial\varphi^- -b^{-i}_{L,R}A^i_{L,R})
(\bar\partial\varphi^- -b^{-i}_{L,R}\bar A^i_{L,R}) \nn
&& +2\cos 2l \{\partial\varphi^+ + \partial\varphi^- 
-(b^{+i}_{L,R}+b^{-i}_{L,R})A^i_{L,R}\} \nn
&& \hskip 1cm \times 
\{\bar\partial\varphi^+ - \bar\partial\varphi^- 
-(b^{+i}_{L,R}-b^{-i}_{L,R})\bar A^i_{L,R}\} \nn
&& +4(\partial l- a^{+i}_{L,R} A^i_{L,R})
(\bar \partial l - a^{+i}_{L,R}\bar A^i_{L,R})
\Bigr]
\eea
The constraint term is also added,
\be
\label{constraint2}
\cL^{\rm constraint}={2k \over \pi} \theta^i 
(\partial \bar A^i - \bar \partial A^i 
+ 2 \epsilon^{ijk} A^j \bar A^k) .
\ee
If there is not the constraint term, the action where 
only left-handed or right-handed $SU(2)$ symmetry is 
gauged will leads to an inconsistent string 
theory since the condition $L_0 - \bar L_0 =0$ would 
not be satisfied. 
However the gauged theory treated here is equivalent 
to the original theory due to the constraint term 
(\ref{constraint2}).\footnote{
When we regard Wess-Zumino-Witten model as a free fermion 
system, there appears an anomaly term proportional to 
$\tr A\bar A$ when the current is coupled with gauge field.  
The anomaly term is already included in the action (\ref{gLWZW})
of the Wess-Zumino-Witten model at the classical level.}
By choosing a gauge condition 
\be
\label{gauge2}
g=\e^{{i \over 4}\pi\sigma^1}\ ,
\ee
(The reason why we do not choose the condition $g=1$ is 
that the Lagrangean (\ref{gLWZW}) contains a term proportional 
to $\cot l$, which is singular at $l=0$.) 
and integrating the gauge fields, we obtain the Lagrangean 
dual to (\ref{LWZW}),
\bea
\label{dualWZW}
\cL^{\rm dual}&=&{k \over \pi}{4 \over 1 + 4 (\theta)^2}
\left[\delta_{ij}-2\epsilon_{ijk}\theta^k +4\theta^i\theta^j
\right]\partial\theta^i\bar\partial\theta^j \nn
-{1 \over 4\pi}R^{(2)}\ln (1 + 4(\theta)^2) \ .
\eea
Note that the Lagrangean (\ref{dualWZW}) is identical with 
that of the chiral model (\ref{dualchi}) accompanying the 
dilaton term (\ref{dilaton}) when
\be
\label{coupling}
e^2={\pi \over k}\ .
\ee
The equivalence might tell that the Wess-Zumino term would 
be induced 
when the chiral model couples with gravity.
The coupling constant $e^2$ in the chiral model does not 
satisfy Eq.(\ref{coupling}) in general.
We expect that the coupling constant $e^2$ would be 
renormalized to satisfy Eq.(\ref{coupling}) due to 
the effect of the gravity since the 
$\beta$-function vanishes in the Wess-Zumino-Witten model 
when the coupling constant satisfies 
Eq.(\ref{coupling}), {\it i.e.}, Eq.(\ref{coupling}) 
corresponds to the fixed point of the renormalization 
group \cite{WZW}. 

\ 

\noindent
{\bf Acknowledgement}

The author would like to thank A. Sugamoto for discussions. 
He is also indebted to K. Fujii who called his attention to 
non-abelian $T$-duality.

\end{document}